%% file: viral_pna_risk.tex
\documentclass[11pt]{article}
\pdfoutput=1
\usepackage{longtable}
\usepackage[colorinlistoftodos]{todonotes}
\usepackage[frozencache,cachedir=.]{minted}
\usepackage{gensymb}
\usepackage{graphicx}
\usepackage{subcaption}
\usepackage{amssymb}

\usepackage{chngcntr}
\counterwithin{figure}{section}

\usepackage[symbol]{footmisc}

\makeatletter
\newcommand{\printfnsymbol}[1]{
  \textsuperscript{\@fnsymbol{#1}}
}
\makeatother
\usepackage{natbib}
\usepackage[margin=1in]{geometry}
\usepackage{libertine}
\usepackage[T1]{fontenc}
\usepackage{hyperref}
\hypersetup{colorlinks,citecolor=blue,urlcolor=blue,linkcolor=blue,linktocpage=true}

\newcommand*\samethanks[1][\value{footnote}]{\footnotemark[#1]}

\begin{document}
\title{Predicting Mortality Risk in Viral and Unspecified Pneumonia 
to Assist Clinicians with COVID-19 ECMO Planning
}

\author{Helen Zhou\thanks{Equal contribution.},
Cheng Cheng\samethanks, Zachary C. Lipton,
George H.~Chen,
Jeremy C. Weiss\\
Machine Learning Department, Heinz College, \\
Carnegie Mellon University \\
\small{\texttt{\{\href{mailto:hlzhou@cmu.edu}{hlzhou},\href{mailto:ccheng2@cmu.edu}{ccheng2},\href{mailto:zlipton@cmu.edu}{zlipton},\href{mailto:georgechen@andrew.cmu.edu}{georgechen},\href{mailto:jeremyweiss@andrew.cmu.edu}{jeremyweiss}\}@cmu.edu
}}}

\date{}
\maketitle

\vspace{-1em}
\begin{abstract}
Respiratory complications due to coronavirus disease COVID-19
have claimed tens of thousands of lives in 2020. 
Many cases of COVID-19 escalate from Severe Acute Respiratory Syndrome (SARS-CoV-2) 
to viral pneumonia to acute respiratory distress syndrome (ARDS) to death. 
Extracorporeal membranous oxygenation (ECMO) is 
a life-sustaining oxygenation and ventilation therapy 
that may be used for patients with severe ARDS
when mechanical ventilation is insufficient to sustain life. 
While early planning and surgical cannulation for ECMO can increase survival,
clinicians report the lack of a risk score hinders these efforts.
In this work, we leverage machine learning techniques
to develop the PEER score, used to highlight critically ill patients 
with viral or unspecified pneumonia 
at high risk of mortality or decompensation 
in a subpopulation eligible for ECMO. 
The PEER score is validated on two large, 
publicly available critical care databases
and predicts mortality at least as well as other existing risk scores. 
Stratifying our cohorts into low-risk and high-risk groups, 
we find that the high-risk group also has a higher proportion 
of decompensation indicators such as vasopressor and ventilator use. 
Finally, the PEER score is provided in the form of a nomogram 
for direct calculation of patient risk, and can be used to highlight 
at-risk patients among critical care patients eligible for ECMO.
\end{abstract}

\section{Introduction}\label{sec:intro}
Coronavirus disease COVID-19 has spread globally,
resulting in millions of infections.
Many cases of COVID-19 progress from Severe Acute Respiratory Syndrome (SARS-CoV-2)
to viral pneumonia to acute respiratory distress syndrome (ARDS) to death.
Extracorporeal membrane oxygenation (ECMO) can temporarily sustain patients with severe ARDS
to bridge periods of time when oxygenation via the lungs 
cannot be achieved via mechanical ventilation. 
However, ECMO is expensive and applicable only for patients
healthy enough to recover and return to a high functional status. 
While ECMO is more effective when planned in advance~\citep{combes2018extracorporeal},
applicable risk scores remain unavailable~\citep{liang2020handbook,acchub}. 

This paper introduces the Viral or Unspecified \textbf{P}neumonia \textbf{E}CMO-\textbf{E}ligible \textbf{R}isk Score, or PEER score for short. This risk score focuses on a critical care population eligible for ECMO and uses measurements acquired at the time of would-be planning---early in the critical care stay. 
While other pneumonia-based risk scores exist~\citep{psiport,Marti2012,smartcop,lim2003defining}, 
our PEER score specifically addresses risks associated with 
viral or unspecified pneumonia in the critical care setting
and the cohort potentially eligible for ECMO. 
Inclusion of unspecified pneumonia broadens the population studied, 
and is chosen because the infectious etiology of pneumonia often cannot be determined.

Though limited by geographic availability,
ECMO usage has increased 4-fold in the last decade~\citep{Ramanathan2020}.
COVID-19 guidelines suggest ECMO as a late option in the escalation of care 
in severe ARDS secondary to SARS-CoV-2 infection~\citep{liang2020handbook,alhazzani2020surviving}.
However, early evidence from epidemiological studies 
of coronavirus~\citep{Wang2020,Yang2020,zhou2020clinical} 
has been insufficient to establish ECMO's utility.
A pooled analysis of four studies~\citep{henry2020covid} 
showed mortality rates of 95\% with the use of ECMO vs. 70\% without,
but the number of patients receiving ECMO was small,
and none of these studies were controlled
nor did they specify indications for~ECMO.

In this paper, to better understand the role of ECMO as a rescue 
for ventilation non-responsive, SARS-CoV-2 ARDS, 
we study risk of death in ARDS among ECMO-eligible patients.
Treatment guidelines for ARDS suggest ECMO use in severe ARDS 
alongside other advanced ventilation strategies~\citep{matthay2020treatment}.  
World Health Organization interim guidance also suggests ECMO for ARDS~\citep{world2020clinical}, 
citing effectiveness in ARDS and mortality reductions in Middle East Respiratory Syndrome (MERS).
Despite these recommendations and the resources allocated 
to ECMO for these situations~\citep{Ramanathan2020}, 
risk scores tailored to ECMO consideration are lacking. 
Our study addresses this need by drawing 
from viral and source unidentified cases of pneumonia 
that escalate to critical care admissions,
guided by the intuition that ARDS associated with these pneumonia
are expected to better resemble COVID-19 ARDS than all-comer ARDS.

We develop and evaluate a risk score for critical care patients 
with viral or unspecified pneumonia and who are eligible for ECMO. 
Our risk score is validated on two large, 
publicly available critical care databases
and predicts mortality at least as well as other existing risk scores. 
We stratify our cohorts into low-risk and high-risk groups, 
and find that the high-risk group also has a higher proportion 
of decompensation indicators such as vasopressor and ventilator use. 
Finally, we provide the risk score in the form of a nomogram 
for direct calculation of patient risk.
We provide this risk score as an assistive tool to highlight at-risk patients 
among critical care patients eligible for ECMO.

\subsection*{Related Work}
There are a number of pneumonia~\citep{curb65,psiport,smartcop,lim2003defining,Guo2019}, COVID-19~\citep{maastrichtE6,Jiang2020,Gong2020}, 
hospitalization mortality~\citep{zimmerman2006acute}, 
and ECMO risk scores~\citep{Schmidt2015}, 
but none are focused around the time of risk evaluation for ECMO candidacy.
The pneumonia and COVID-19 risk scores are assessed on populations with lower acuity, 
while APACHE is not focused on patients with respiratory illness.
Our risk score focuses at the stage of ECMO planning 
rather than among patients already receiving ECMO.

Registry-based studies have also considered expected mortality variation in SARS-CoV-2 
compared to that of other viral infections, including MERS, H1N1 flu, and seasonal flu. 
One MERS-related ARDS study of critically ill patients demonstrated 
a higher mortality rate than those in studies on COVID-related ARDS,  
but may be attributed to more severely ill patients at enrollment~\citep{arabi2017critically}. 
A similar study of H1N1 reported lower mortality rates (12-17\%), 
albeit considering a younger population with average age of 40~\citep{Aokage2015}.

Physiologic concerns have also been raised about the use of ECMO for SARS-CoV-2. 
While ECMO is primarily beneficial for respiratory recovery, 
some argue that a spike in all-cause death but not ARDS-related death 
could indicate a limited role of ECMO~\citep{henry2020poor}.
In addition, COVID-associated lymphopenia might be exacerbated 
by ECMO-induced lymphopenia which could mechanistically affect a healthy immune response to infection.  
Inflammatory cytokines and specifically interleukin 6 elevation 
is associated with COVID-19 mortality and
rises with the use of ECMO~\citep{henry2020covid,bizzarro2011infections}. 
These expert voices do not argue for the avoidance of ECMO, 
but rather call for additional study.

\section{Data}

The eICU Collaborative Research Database~\citep{eicu} contains data 
for 200,859 admissions to intensive care units (ICU) 
across multiple centers in the United States between 2014 and 2015.
The MIMIC-III clinical database~\citep{mimiciiidata} 
consists of data from 46,476 patients who stayed in critical care units 
of the Beth Israel Deaconess Medical Center between 2001 and 2012.
Our analysis and model development is done primarily using data 
derived from eICU, and externally validated on data derived from MIMIC-III.

\begin{figure}[H]
\centering
\begin{subfigure}[t]{0.45\textwidth}
\includegraphics[width=0.85 \textwidth]{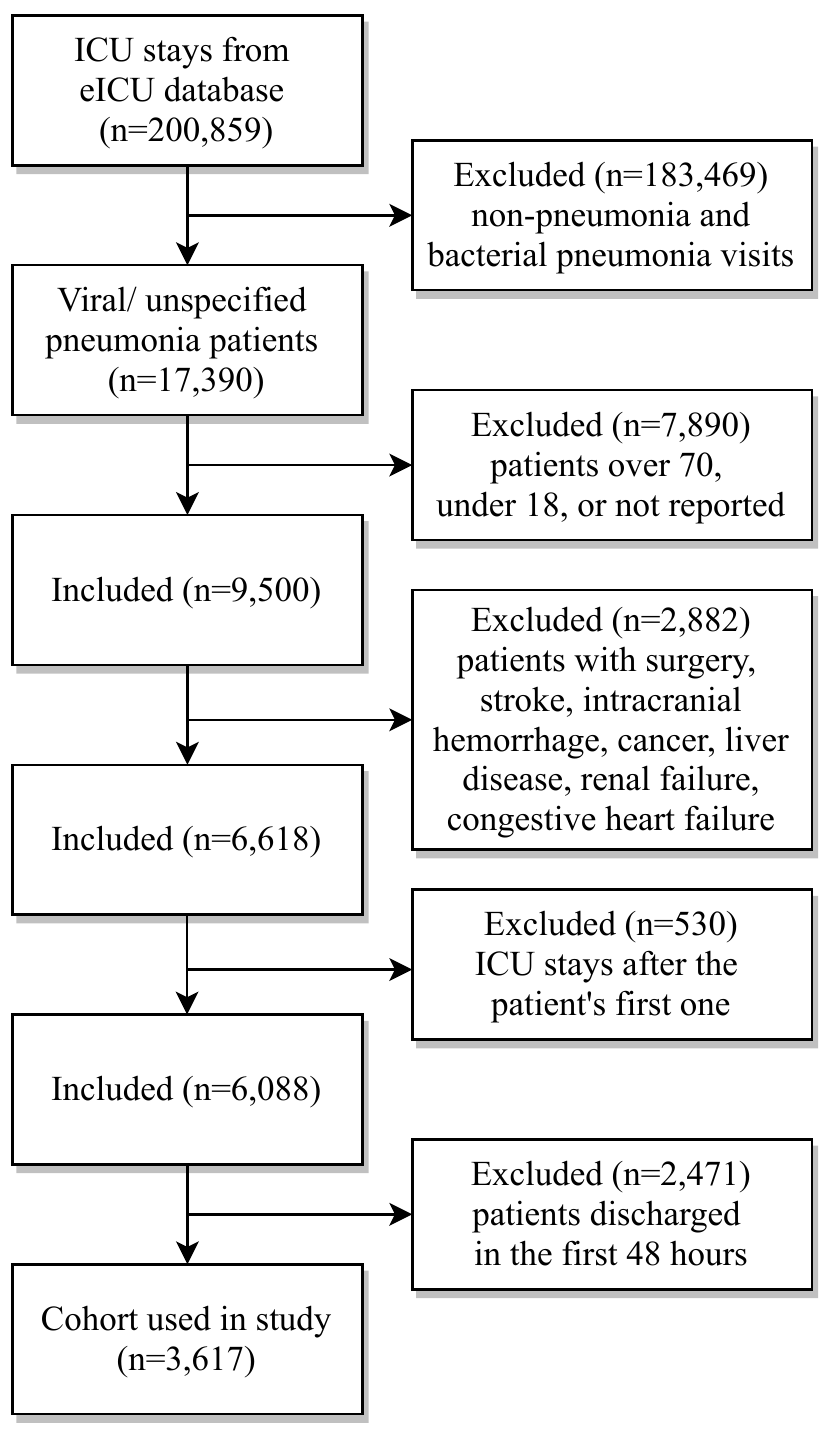}
\caption{eICU cohort selection}
\label{fig:eicu_cohort_selection}
\end{subfigure}
\hspace{1em}
\begin{subfigure}[t]{0.45\textwidth}
\includegraphics[width=0.85 \textwidth]{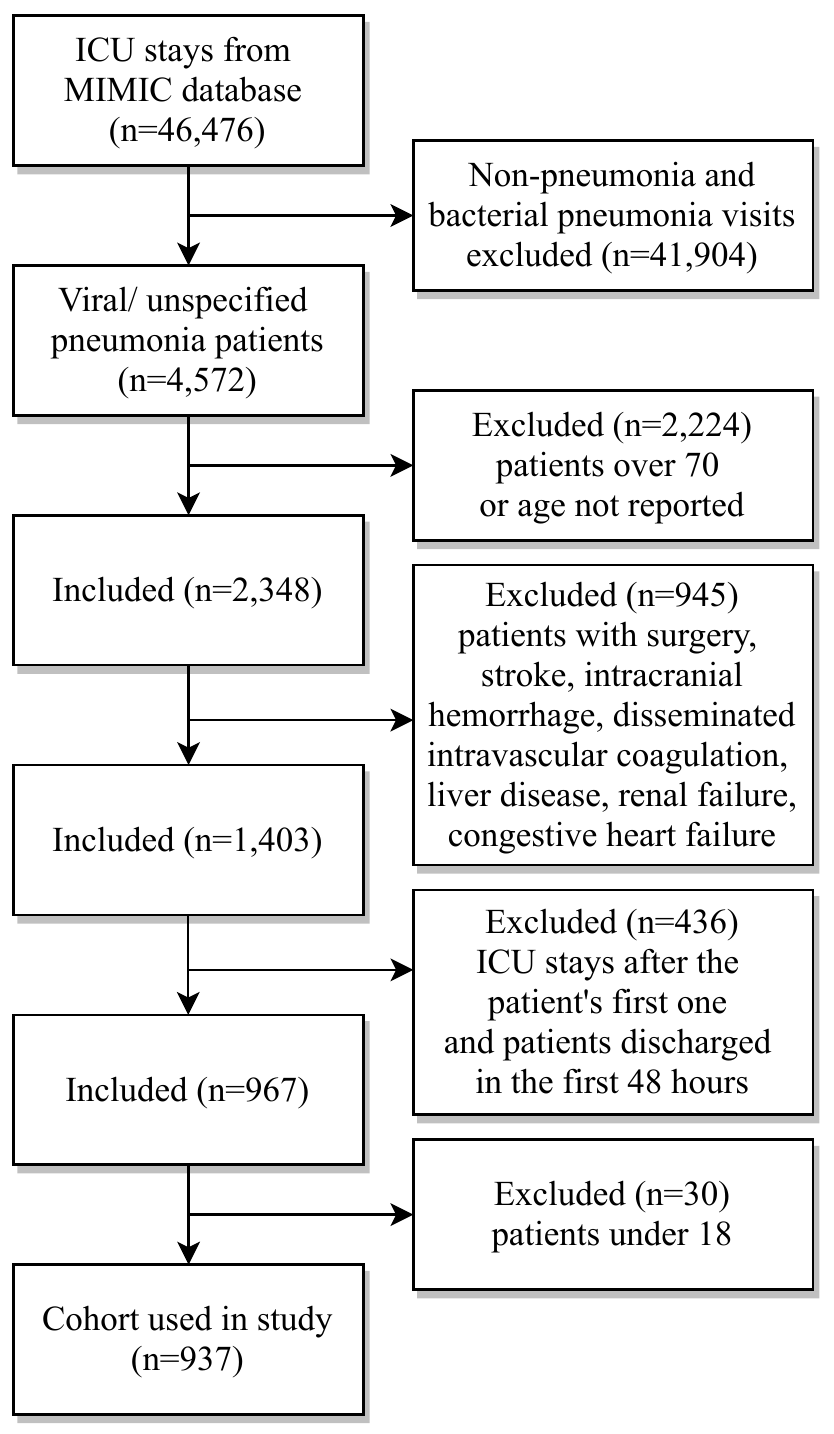}
\caption{MIMIC-III cohort selection}
\label{fig:mimic_cohort_selection}
\end{subfigure}
\vspace{-0.5em}
\caption{Inclusion and exclusion criteria for cohorts extracted from eICU and MIMIC. 
The disseminated intravascular coagulation exclusion criteria
could not be reliably extracted from eICU due to missing data.}
\label{fig:flowdiagrams}
\vspace{-0.5em}
\end{figure}

\vspace{1em}\noindent\textbf{Cohort Selection}~
Data for the study cohort were extracted according to 
the inclusion and exclusion criteria summarized in Figure \ref{fig:flowdiagrams}.
Our population of interest is patients with viral or otherwise unspecified 
non-bacterial, non-fungal, non-parasitic, and non-genetic pneumonia. 
While there are no absolute contraindications of ECMO,
the therapy is reserved for patients likely to have functional recovery. 
Thus, patients over 70 years old were excluded 
since they would not be good candidates for ECMO,
and patients under 18 were excluded because
SARS-CoV-2 pneumonia progressing to hypoxic respiratory failure 
is exceedingly rare in this age group.
Other relative contraindications to ECMO include 
stroke, intracranial hemorrhage, disseminated intravascular coagulation, cancer, 
liver disease, renal failure, patients undergoing surgery, and congestive heart failure. 
Over the course of each patient's hospital stay, 
the first ICU stay was analyzed. 
Patients who died or were discharged within the first 48 hours of being admitted 
were excluded to focus on the stage of critical care after initial entry 
when lower-risk oxygen supplementation strategies
(\emph{e.g.}, ventilation) are being performed,
and, methodologically, to provide a richer set of features for prediction. 
Table \ref{tab:cohort_table_full} summarizes demographics and outcomes of the cohorts.

\begin{table}[H]
\caption{Characteristics and treatment of patients with viral or unspecified pneumonia in eICU and MIMIC-III cohorts.
Data are median (Q1-Q3) or count (\% out of n).}
\vspace{-0.25em}
  \setlength{\tabcolsep}{2.5em}
  \resizebox{0.95\textwidth}{!}{\begin{minipage}{\textwidth}
    \centering
    \scriptsize
    \include{tables/cohort_table}
  \end{minipage}}
  \label{tab:cohort_table_full}
\end{table}

\vspace{0.15em}\noindent\textbf{Data Extraction}~
The eICU and MIMIC cohorts were extracted using string matching on diagnosis codes 
and subsequent clinician review of the descriptors
(provided in Appendix \ref{appendix:cohort_selection}). 
Features were manually merged from several tables into their corresponding elements 
based on a process of visualization, query, and physician review. 
This involved harmonizing feature units (e.g. Farenheit to Celsius), 
removing impossible values (e.g. negative blood pressures), 
and merging redundant data fields 
(e.g., filling in the sum of the individual Motor, Verbal, and Eye Opening Score fields 
for the Glasgow Coma Score Total when it was missing). 
We provide additional feature and outcome extraction details 
in Appendix \ref{appendix:feature_outcome_extraction}.

All features were combined into a fixed-length vector,
using the most recent value prior to 48 hours after ICU admission. 
Before imputation, approximately half of the features had missingness below 5\%, and 80\% of the features 
had missingness below 30\%, however several variables have high missingness (Appendix \ref{appendix:missingness}). 
Missing values were imputed using MissForest~\citep{missforest} 
and our model was not sensitive to MissForest imputation 
(Appendix \ref{appendix:missforest_sensitivity}).

\vspace{1em}\noindent\textbf{Features}~
Features were extracted from demographics, comorbidities, vitals, 
physical exam findings, and laboratory findings routinely collected in critical care settings.
Numerical features were normalized, and categorical features 
were converted into binary features (one less than the number of categories). 
The 52 features provided to the model are listed in Appendix \ref{appendix:feature_outcome_extraction}. 

\vspace{1em}\noindent\textbf{Outcomes}~ Our primary outcome of interest is in-ICU mortality. 
Secondary outcomes indicating decompensation include 
(1) vasopressor use and (2) mechanical ventilation use. 
For each outcome, we define the time to event 
as the time to first outcome or censorship, where censorship corresponds to discharge from the ICU. 
For details, see Appendix \ref{appendix:feature_outcome_extraction}.

\section{Methods}
\label{sec:methods}
\noindent\textbf{Lasso-Cox}~
To predict patients' survival time, we use the Cox proportional hazards model~\citep{coxph} 
with L1 regularization, referred to as \emph{Lasso-Cox}~\citep{Rcoxnet}. 
We choose Lasso-Cox for its ease of interpretation and calculation, 
owing to its selection of sparse models.\footnote{We also tried the Cox model 
with elastic-net regularization (combined L1 and L2 regularization) 
but found little to no gain in concordance.} 
In particular, for a patient with covariates $\textbf{x}\in\mathbb{R}^d$, 
the predicted log hazard of the patient is $\beta^{\top} \textbf{x}$ 
(higher log hazard is associated with shorter survival time), 
where $\beta\in\mathbb{R}^d$ is the vector of Cox regression coefficients 
that can be interpreted as log hazard ratios. 
During model fitting, L1 regularization $\lambda\sum_{j=1}^d |\beta_j|$ is used 
to encourage only a few features to have nonzero $\beta$ value, 
where $\lambda>0$ is a user-specified hyperparameter.

\vspace{1em}\noindent\textbf{Evaluation Metrics}~To evaluate model performance,
we consider concordance and calibration. 
\emph{Concordance} (c-index) is a common measure
of goodness-of-fit in survival models~\citep{harrellconcordance}, 
defined as the fraction of pairs of subjects whose survival times 
are correctly ordered by a prediction algorithm, among all pairs that can be ordered.

To compute concordance confidence intervals we use 1000 bootstrapped replicates, 
each the size of the dataset and sampled with replacement.
\emph{Calibration} is evaluated by plotting 
the Kaplan-Meier observed survival probability 
versus the predicted survival probability. 
Using the R package hdnom~\citep{hdnom}, we construct 
our calibration plots (Figure \ref{fig:calibration}) 
with 1000 bootstrap resamplings for internal calibration, 
using $5$ groups and a prediction time of $3$ days.

\vspace{1em}\noindent\textbf{Experimental Setup}~
We divided the eICU cohort into a training set (70\% of the data, n=2537) 
and test set (30\%, n=1080). 
The eICU training set is used for model selection and analysis,
whereas the eICU test set and entirety of the MIMIC cohort are used for model evaluation.
Throughout our analysis, we compare our risk score (PEER) 
to three pneumonia risk scores: CURB-65~\citep{curb65}, 
PSI/PORT~\citep{psiport}, and SMART-COP~\citep{smartcop}; 
and one COVID-19 risk score: GOQ~\citep{maastrichtE6}.

\vspace{1em}\noindent\textbf{Model selection}~
We select $\lambda$ via 10-fold cross validation 
and grid search on the eICU training set
to maximize concordance subject to sufficient sparsity. 
We observe that $\lambda=0.01$ gives the best trade-off between concordance ($0.73$) 
and number of features selected ($18$), 
as a $0.01$ increase in concordance corresponds to 10 additional non-zero features.
To check the stability of this hyperparameter choice, 
we imputed our data using ten random seeds and ran 10-fold cross validation on the resulting datasets. 
Across all runs, $\lambda=0.01$ achieved concordance of approximately $0.73$ 
and selected similar features and coefficients. 
Additional details about grid search, the concordance and sparsity tradeoff, 
and robust selection of coefficients can be found in Appendix \ref{sec:lambda}, 
Appendix Figure \ref{fig:missforest10_concordance}, 
and Appendix Figure \ref{fig:missforest10_coefficients}. 

\vspace{1em}\noindent Code for data extraction and all model results is available at \href{https://github.com/hlzhou/peer-score}{https://github.com/hlzhou/peer-score}.

\section{Results}

\begin{figure}[t]
    \centering
    \includegraphics[width=0.7\textwidth]{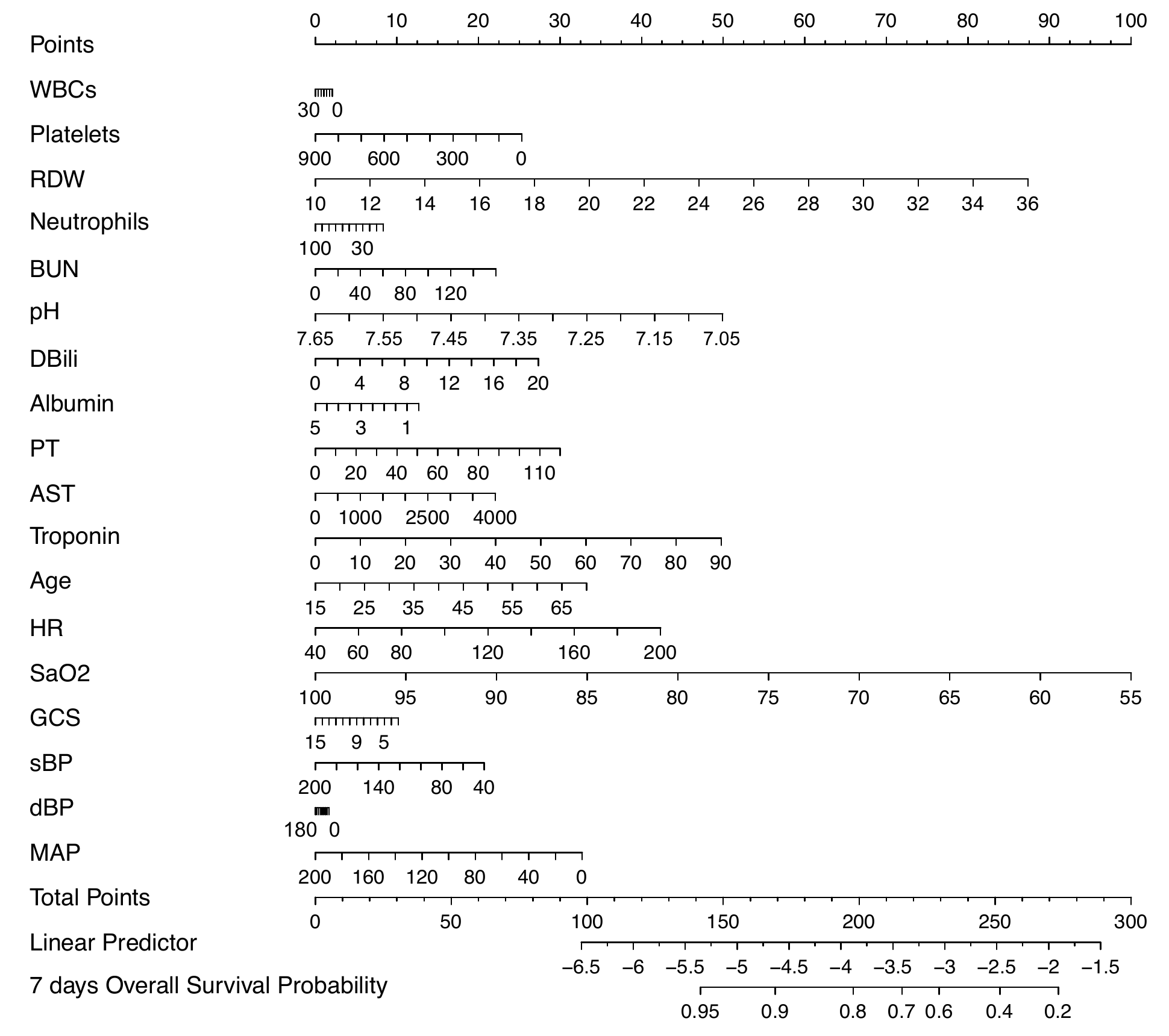}
    \vspace{-0.5em}
    \caption{Nomogram for manual calculation of the PEER score.}
    \label{fig:nomogram}
    \vspace{-0.75em}
\end{figure}

As explained in Section \ref{sec:methods},
model selection yielded a L1-regularized Cox model with $\lambda = 0.01$. 
The learned hazard ratios, i.e. the PEER score, for normalized data are displayed in Table \ref{tab:betas} 
and as a boxplot in Appendix Figure \ref{fig:betasboxplot}. 
For reference, Appendix \ref{appendix:betasboxplot} contains 
the standard deviation and mean of each feature in the model. 
To assist with easy risk score calculation,
we also provide a nomogram~\citep{hdnom} in Figure \ref{fig:nomogram}\footnote{To use this nomogram, look up a patient's original (pre-normalization) values, match it to a number of points listed across the top, and look up the sum of that patient's points in the scale across the bottom.}.

\begin{table}[H] 
    \caption{Hazard ratios for the Lasso-Cox model, i.e. the PEER score, excluding hazard ratios equal to 1 (since they do not contribute to the model). Hazards ratios (HR) and 95\% confidence intervals (CI) are reported on normalized~data.}
  \setlength{\tabcolsep}{2.5em}
  \resizebox{\textwidth}{!}{\begin{minipage}{\textwidth}
    \centering
    \include{tables/betas}
    \label{tab:betas}
  \end{minipage}}
\end{table}

\begin{table}[t]
\caption{Concordances of the PEER score, CURB-65, PSI/PORT, SMART-COP, and GOQ. Bootstrapping with 1000 replicates was used to compute 95\% confidence intervals (in parentheses).}
  \setlength{\tabcolsep}{5pt}
  \resizebox{\textwidth}{!}{\begin{minipage}{\textwidth}
    \centering
    \include{tables/concordances}
  \end{minipage}}
  \vspace{0.5em}
  \label{tab:concordances}
\end{table}

To evaluate discriminative ability, we use concordance.
On both the eICU and MIMIC datasets, our Lasso-Cox model (which we name the PEER score) achieves concordance 
greater than or comparable to that of existing risk scores (Table \ref{tab:concordances}). 
On the eICU test set, the PEER score achieves the highest concordance among the risk scores, $0.77$, 
with the second highest, $0.73$, attained by SMART-COP. 
On the MIMIC dataset, PEER and SMART-COP again achieve the highest concordances, 
but concordance degrades to $0.66$ for both scores. 
One possible reason for this degradation is that arterial oxygen saturation (SaO2) 
is a significant feature and only $1.5\%$ of these values are missing in eICU, 
but in MIMIC $72.6\%$ of these values are missing (Appendix~\ref{appendix:missingness}).

\begin{figure}[H] 
\centering
\begin{subfigure}{0.33\textwidth}
    \includegraphics[width=0.9\textwidth]{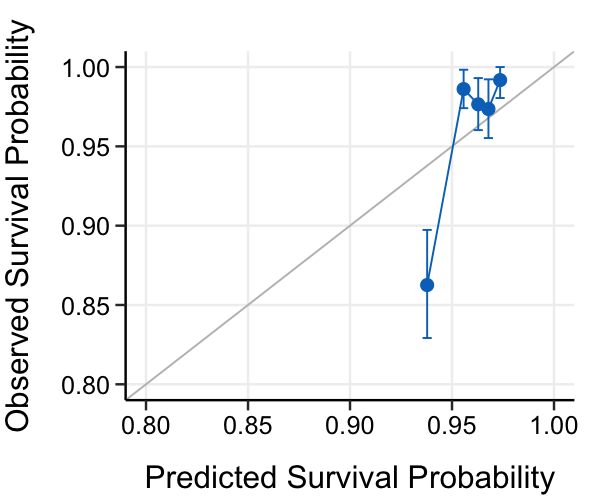}%
    \caption{Train eICU}
    \label{fig:cal_train}
\end{subfigure}%
\begin{subfigure}{0.33\textwidth}
    \includegraphics[width=0.9\textwidth]{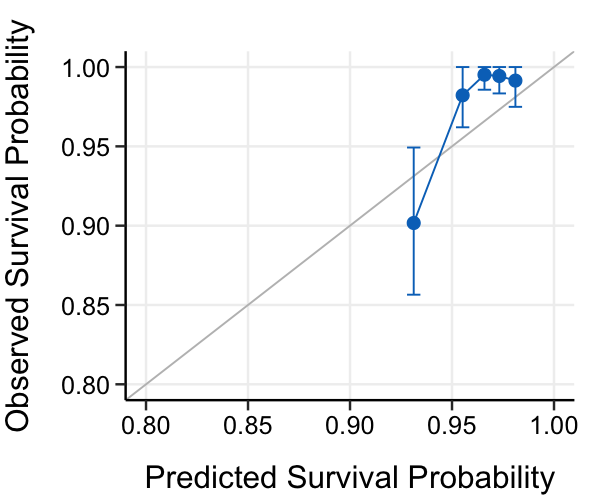}%
    \caption{Test eICU}
    \label{fig:cal_test}
\end{subfigure}%
\begin{subfigure}{0.33\textwidth}
    \includegraphics[width=0.9\textwidth]{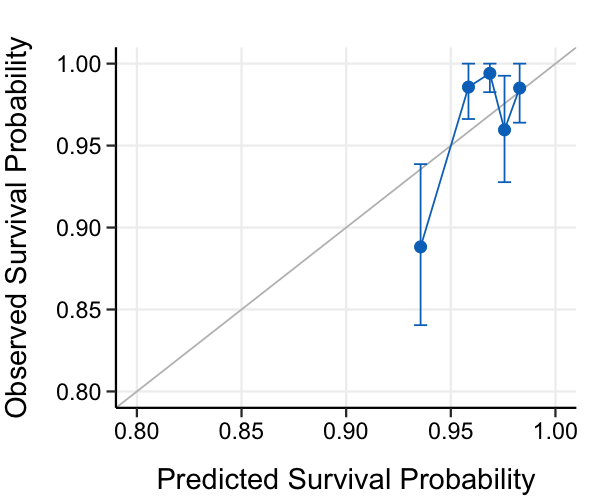}%
    \caption{MIMIC}
    \label{fig:cal_mimic}
\end{subfigure}
\caption{Calibration plots with 95\% confidence intervals on train eICU, test eICU, and test MIMIC.}
\label{fig:calibration}
\end{figure}

From the PEER score calibration curves (Figure \ref{fig:calibration}),
we see one high risk group separate from the other low risk groups. 
While predicted survival probability of the high risk group is overestimated in the training set, 
it is estimated within confidence intervals in both test sets.

\begin{figure}[H]
    \centering
    \includegraphics[width=\linewidth]{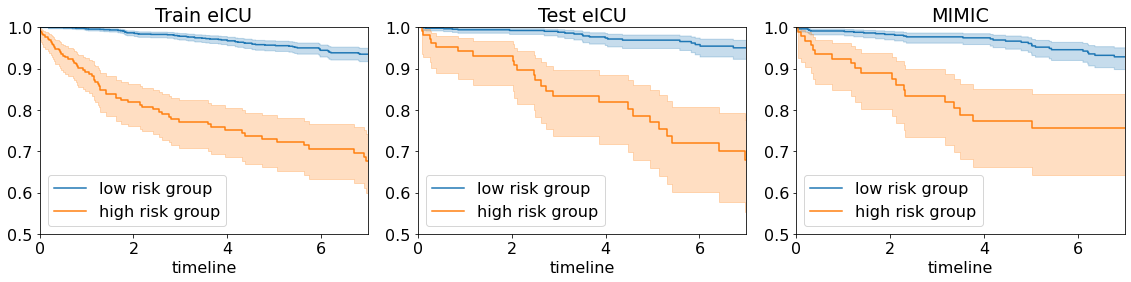}
    \vspace{-2em}
    \caption{Kaplan-Meier survival curves of high vs.~low risk groups in train eICU, test eICU, and MIMIC. Shaded regions are the 95\% confidence intervals.}
    \label{fig:surv}
    \vspace{-1em}
\end{figure}

To compare low-risk and high-risk populations (as defined by our model), 
we set a threshold at the 90th percentile of the predicted risk value from the training set.
We find that over a 7 day time period, 
the Kaplan-Meier survival curves for the high and low risk subpopulations 
are clearly distinct Figure~\ref{fig:surv})~\citep{py_lifelines}.
At 7 days, the high risk and low risk groups in the eICU test set 
have survival proportions $0.68$ and $0.95$, respectively. 
The gap is smaller in MIMIC with survival probabilities of $0.75$ and $0.95$, respectively. 
In each case, the difference in survival is larger for the PEER score
than for other risk scores (Appendix Figure~\ref{fig:allscores_strata}, 
with stratification details in Appendix \ref{appendix:survcurves}). 

\begin{figure}[H] 
\begin{subfigure}{.5\textwidth}
\centering
\includegraphics[width=0.7\textwidth]{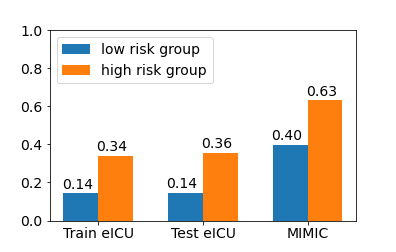}
\caption{vasopressor}
\label{fig:vaso}
\end{subfigure}
\begin{subfigure}{.5\textwidth}
\centering
\includegraphics[width=0.7\textwidth]{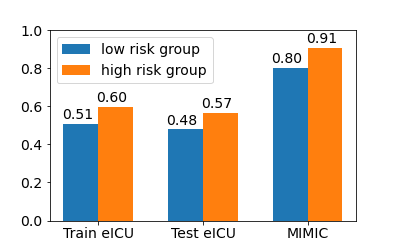}
\caption{ventilator}
\label{fig:vent}
\end{subfigure}
\caption{Proportion of high and low risk patients who received vasopressors or ventilators. 
High and low risk groups are derived from the PEER score.}
\label{fig:secondary}
\end{figure}

In addition to mortality, decompensation can be indicated by vasopressor and ventilator use. 
As expected, vasopressors and ventilators are more commonly used in the high risk group (Figure \ref{fig:secondary}).

\section{Discussion} 

Our PEER score achieves greater or comparable concordance to baselines 
on the eICU (in-domain) and MIMIC (out-of-domain) test sets. The Lasso-Cox model selects 18 features, 
making it easy to calculate. 
We find that lower SaO2, which is associated with poorer oxygenation status,
is predictive of decompensation.  
As expected, we also find that old age is predictive of death.
Red blood cell distribution width, associated with expanded release of immature red blood cells 
as a response to insufficient oxygen delivery to tissues, 
is also a strong risk factor for death in patients with COVID-19~\citep{Gong2020}. 
Three variables (pH, prothrombin time, and age) violate 
the proportional hazards assumption of the Cox model, 
and Lasso-Cox models shrink coefficients towards $0$,
so the hazard ratios themselves should be interpreted with caution. 

Stratifying each cohort into low risk and high risk subpopulations based the PEER score, 
there is a clear separation in their survival curves (Figure \ref{fig:surv}) across all three datasets, 
with the low risk group observing a survival of 0.9 or greater after 7 days 
and the high risk group observing a survival of 0.75 or less. 
The mortality-based high risk groups also are associated with our proximate indicators of decompensation: 
vasopressor and ventilator use (Figure \ref{fig:secondary}). 
Compared to the survival curves for low and high risk subpopulations derived from other risk scores, 
the survival curves from the PEER score are the most distinctly separated (Appendix \ref{appendix:survcurves} Figure \ref{fig:allscores_strata}).

Calibration plots for the PEER score also show a high risk group 
clearly separated from the rest (Figure \ref{fig:calibration}). 
While the survival probability of the high risk group 
is overestimated in the training set, 
it is within error bars in the test set. 

For ECMO allocation, practically, accurate \emph{ranking} of risk, as measured by concordance, 
may be more important than the probabilities predicted. 
While the PEER score outperformed other risk scores in the eICU test set, 
there was a decline in performance on the MIMIC test set, 
and the performance of the PEER score was comparable to that of SMART-COP.
We also note that in MIMIC, arterial oxygen saturation (SaO2) has $72.6\%$ missingness 
yet is an important feature of the PEER score. 
In contrast, it has $1.5\%$ missingness in eICU. 
This demonstrates the importance of thinking critically about how our risk score, 
which was trained on the eICU cohort and depends on $18$ specific features, 
generalizes to the population to which the score is being applied.

\vspace{1em}
\noindent\textbf{Limitations and Future Work}~
Importantly our cohort is defined not by COVID-19 positive pneumonia patients 
but instead by viral or unspecified pneumonia patients who are ECMO-eligible. 
While our risk score demonstrates good discriminative ability and is interpretable, 
there are several additional decision-making considerations beyond the scope of this paper. 
Clinicians interested in applying the risk score to COVID-19 pneumonia should consider 
how representative this population is of their own. 
Because ECMO is a constrained resource, there are also ethical questions about who should get treatment. 
This risk score does not attempt to address these questions, 
but simply provides relevant information to those making such decisions. 
More broadly, we hope to provide this risk score as a potential resource 
for future SARS-like diseases that require ECMO consideration. 

\clearpage
\bibliographystyle{abbrvnat}
\bibliography{viral_pna_risk}

\clearpage
\appendix
\section{Cohort Selection} \label{appendix:cohort_selection}
The cohort for model development and analysis on a held-out set was extracted 
from the eICU Collaborative Research Database Version 2.0. 
A cohort was also extracted from MIMIC-III Version 1.4 for external validation.

For both the eICU and MIMIC cohorts, viral or unspecified pneumonia patients 
were included by string matching of International Classification of Disease (ICD 9) diagnosis code descriptions
and subsequent clinician review of these descriptors. 
Using the patient and demographics tables, patients over 70 or under 18 were excluded.
Other relative contraindications to ECMO were likewise excluded 
by string matching and clinician review of the resulting descriptors. 

\subsection{eICU Cohort} 
Below are the \verb diagnosisstring 's used to select patients with viral or unspecified pneumonia.

\begin{scriptsize}
\begin{verbatim}
    'infectious diseases|chest/pulmonary infections|pneumonia|ventilator-associated',
    'surgery|respiratory failure|ARDS|pulmonary etiology|pneumonia',
    'infectious diseases|chest/pulmonary infections|pneumonia|community-acquired|viral',
    'infectious diseases|chest/pulmonary infections|pneumonia',
    'infectious diseases|chest/pulmonary infections|lung abscess|secondary to pneumonia',
    'pulmonary|pulmonary infections|pneumonia|community-acquired|viral|respiratory syncytial',
    'pulmonary|pulmonary infections|pneumonia|community-acquired',
    'surgery|infections|pneumonia|hospital acquired (not ventilator-associated)',
    'infectious diseases|chest/pulmonary infections|empyema|associated with pneumonia',
    'pulmonary|pulmonary infections|pneumonia|hospital acquired (not ventilator-associated)',
    'pulmonary|pulmonary infections|pneumonia|hospital acquired (not ventilator-associated)',
    'infectious diseases|chest/pulmonary infections|pneumonia|opportunistic',
    'surgery|respiratory failure|acute lung injury|pulmonary etiology|pneumonia',
    'pulmonary|respiratory failure|acute lung injury|pulmonary etiology|pneumonia',
    'infectious diseases|chest/pulmonary infections|pneumonia|community-acquired',
    'surgery|infections|pneumonia',
    'pulmonary|pulmonary infections|pneumonia|community-acquired|viral',
    'infectious diseases|chest/pulmonary infections|pneumonia|hospital acquired
        (not ventilator-associated)',
    'pulmonary|respiratory failure|ARDS|pulmonary etiology|pneumonia',
    'pulmonary|pulmonary infections|pneumonia|hospital acquired (not ventilator-associated)
        |viral',
    'transplant|s/p bone marrow transplant|idiopathic pneumonia syndrome - bone marrow 
        transplant',
    'pulmonary|pulmonary infections|lung abscess|secondary to pneumonia',
    'infectious diseases|chest/pulmonary infections|pneumonia|community-acquired|viral|
        respiratory syncytial',
    'pulmonary|pulmonary infections|pneumonia',
    'pulmonary|pulmonary infections|pneumonia|ventilator-associated',
    'pulmonary|pulmonary infections|pneumonia|opportunistic',
    'surgery|infections|pneumonia|community-acquired'
\end{verbatim}
\end{scriptsize}
In addition to filtering for patients between 18-70 years old, 
the following code was used to exclude other contraindications based on ICD codes. 
All resulting descriptors were clinician-reviewed:
\begin{minted}[breaklines, fontsize=\scriptsize]{psql}
CREATE TABLE pna_viral_cohort_exclude AS SELECT DISTINCT d.patientunitstayid
FROM pna_viral_cohort0 AS c
JOIN diagnosis AS d
ON c.patientunitstayid = d.patientunitstayid
WHERE (lower(diagnosisstring) like '%surgery%')
OR (lower(diagnosisstring) like '%neurologic%stroke%')
OR (lower(diagnosisstring) = 'surgery|vascular surgery|surgery-related ischemia|postop stroke')
OR (lower(diagnosisstring) like '%cranial%hemorrhage%')
OR (lower(diagnosisstring) like '%cancer%')
OR (lower(diagnosisstring) like '%tumor%')
OR (lower(diagnosisstring) like '%lymphoma%')
OR ((lower(diagnosisstring) like '%hepatic%' OR lower(diagnosisstring) like '%hepatitis%')
    AND (diagnosisstring NOT IN ('gastrointestinal|post-GI surgery|s/p hepatic surgery',
      'gastrointestinal|hepatic disease|toxic hepatitis',
      'infectious diseases|GI infections|intra-abdominal abscess|hepatic|bacterial',
      'burns/trauma|trauma - abdomen|hepatic trauma',
      'gastrointestinal|abdominal/ general|intra-abdominal abscess|subhepatic',
      'gastrointestinal|hepatic disease|hepatorenal syndrome',
      'gastrointestinal|hepatic disease|hepatic infarction',
      'cardiovascular|shock / hypotension|sepsis|sepsis with single organ dysfunction-acute hepatic failure',
      'toxicology|drug overdose|acetaminophen overdose|hepatic injury expected',
      'gastrointestinal|hepatic disease|hepatic dysfunction|pregnancy related',
      'gastrointestinal|hepatic disease|hepatic dysfunction',
      'gastrointestinal|trauma|hepatic trauma',
      'toxicology|drug overdose|acetaminophen overdose|hepatic injury unexpected')))
OR lower(diagnosisstring) like '%liver disease%'
OR lower(diagnosisstring) like '%congestive heart failure%'
OR ((lower(diagnosisstring) like '%renal%') 
    AND (lower(diagnosisstring) not like '%adrenal%')
    AND (icd9code like '40%'));
\end{minted}

\subsection{MIMIC Cohort}
Here is the R code used to extract the MIMIC cohort:
\begin{minted}[fontsize=\scriptsize]{R}
# exclude based in ICD codes
exclusion = bind_rows(
  fread(paste0(mimicdir,"DIAGNOSES_ICD.csv")) %>% as_tibble() %>%
  inner_join(read_csv(paste0(mimicdir,"D_ICD_DIAGNOSES.csv")) %>% 
             select(ICD9_CODE,LONG_TITLE), by="ICD9_CODE") %>%
  filter(str_detect(str_to_lower(LONG_TITLE), 
         pattern="isseminated intra|cerebral hem|cerebral inf")) %>% 
  filter(!str_detect(str_to_lower(LONG_TITLE), pattern="history")),
  fread(paste0(mimicdir,"DIAGNOSES_ICD.csv")) %>% as_tibble() %>%
    inner_join(read_csv(paste0(mimicdir,"D_ICD_DIAGNOSES.csv")) %>% 
               select(ICD9_CODE,LONG_TITLE), by="ICD9_CODE") %>%
  filter(str_detect(str_to_lower(LONG_TITLE), pattern="urgical")) %>%
  filter(!str_detect(str_to_lower(LONG_TITLE), pattern=" not ")) %>%
  filter(SEQ_NUM < 3) # only remove if surgery in top 3) %>%
  select(SUBJECT_ID) %>% rename(subject_id=SUBJECT_ID) %>% distinct()

exclusion = exclusion %>% bind_rows(
  codx %>%  # codx.tsv list from MIT-LCP MIMIC extraction code
  select(liver_disease, renal_failure,congestive_heart_failure, subject_id) %>% 
  mutate(anyofthem=liver_disease+renal_failure+congestive_heart_failure) %>% 
  filter(anyofthem>0) %>%
  select(subject_id)) %>% distinct()

# Additional filtering. 
# For brevity, the following code only includes the filter operations
# (original code had many joins, mutations, etc. and will released upon publication)
pna_cohort = pna_cohort %>%
   filter(as.numeric(`age at icu admission`)<70) %>%
   filter(! hosp_id %in% exclusion$subject_id) %>%
   filter(censor_or_deceased_days > 0) %>%
   filter(cancer==0)
\end{minted}

\newpage
\section{Features and Outcomes}\label{appendix:feature_outcome_extraction}
Features and outcomes in eICU were extracted from tables containing patient ICU stays, 
demographics, diagnoses, nurse charting, nurse assessments, 
periodic vitals, laboratory findings, and treatment information. 
In MIMIC, this data were extracted from tables describing patient demographics, hospital encounters, ICU encounters, diagnoses, laboratory measurements, 
nurse charting events, event monitoring, and procedures. 

\paragraph{Features} Features were manually merged into their corresponding elements 
based on a process of visualization, query, and physician review. For example, both eICU and MIMIC store the Glasgow Coma Score as a Total and as individual Motor, 
Verbal, and Eye Opening Scores respectively which requires a merge. 

The full list of 52 \textbf{features} our model had access to (prior to model selection):
\begin{minted}{python}
['rbcs', 'wbc', 'platelets', 'hemoglobin', 'hct', 'rdw', 'mcv', 
'mch', 'mchc', 'neutrophils', 'lymphocytes', 'monocytes', 
'eosinophils', 'basophils', 'bun', 'temperature', 'ph', 'sodium', 
'glucose', 'pao2', 'ldh', 'direct_bilirubin', 'total_bilirubin', 
'total_protein', 'albumin', 'pt', 'ptt', 'ast', 'alt', 'creatinine', 
'troponin', 'alkaline_phosphatase', 'bands', 'bicarbonate', 'calcium', 
'chloride', 'potassium', 'age', 'heart_rate', 'sao2', 'gcs', 
'respiratory_rate', 'bp_systolic', 'bp_diastolic', 'bp_mean_arterial',
'pleural_effusion', 'orientation', 'African American', 'Asian', 
'Caucasian', 'Hispanic', 'Male']
\end{minted}

A slightly larger set of features relevant to COVID-19 was initially considered 
(including smoking, C-reactive protein, discharge to nursing home), 
but due to unreliable extraction and high missingness, these were excluded in both our model's training and in calculation of existing risk scores.

\paragraph{Outcomes} Vasopressor use was defined by use of norepinephrine, epinephrine, phenylephrine, vasopressin, milrinone, dobutamine, or dopamine. Mechanical ventilation was defined by the documentation of mechanical ventilator use in the treatment table (eICU) or nurse charting (MIMIC). 
\newpage
\section{Amount of Missing Data in eICU and MIMIC Cohorts}\label{appendix:missingness}
Table \ref{tab:missingness} contains the levels of missingness for the variables in Table \ref{tab:cohort_table_full}.
\begin{table}[H]
\caption{Levels of missing values in each cohort.}
\setlength{\tabcolsep}{2em}
  \resizebox{0.8\textwidth}{!}{\begin{minipage}{\textwidth}
    \include{tables/missingness}
  \end{minipage}}
  \label{tab:missingness}
\end{table}

\section{Grid Search Values for $\lambda$} \label{sec:lambda}
The grid search values for $\lambda$: [0.5, 0.375, 0.25, 0.125, 0.10, 0.075, 0.05, 0.0225, 0.025, 0.0275, 0.02, 0.0175, 0.015, 0.0125, 0.01, 0.005, 0.0005]. Also, Figure \ref{fig:missforest10_coefficients} shows the tradeoff between concordance versus the number of features selected with different $\lambda$.
\section{Nonzero Coefficients in the Model} \label{appendix:betasboxplot}
\begin{figure}[H]
    \centering
    \includegraphics[width=0.7\textwidth]{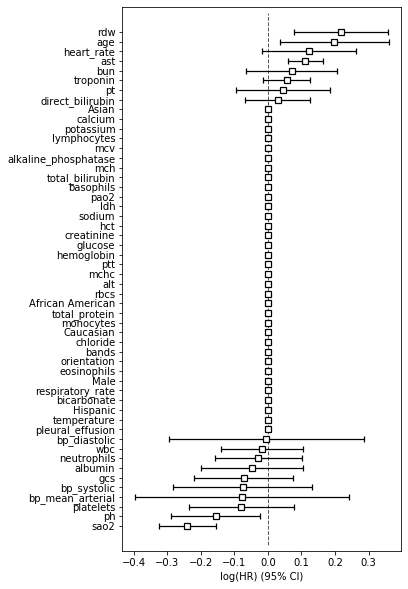}
    \caption{Coefficients of the learned Cox model with penalty 0.01 and L1 regularization, with 95\% confidence intervals (equivalent to what is reported in Table \ref{tab:betas}).}
    \label{fig:betasboxplot}
\end{figure}

Figure \ref{fig:betasboxplot} is a boxplot of the coefficients of the model learned on the normalized data.

For reference, the means and standard deviations of the nonzero features are in Table \ref{tab:mean_std}.
\begin{table}[H]
\caption{Mean and standard deviation of nonzero Lasso-COX features.}
\centering
\setlength{\tabcolsep}{2em}
\include{tables/mean_std}
\label{tab:mean_std}
\end{table}

\section{Sensitivity to MissForest Imputation}\label{appendix:missforest_sensitivity}
MissForest~\citep{missforest} is a non-parametric missing value imputation technique 
which iteratively fits random forests to impute missing values in each column, 
starting with the column with the fewest missing values.
The process repeats itself until the difference between imputed arrays 
over successive iterations meets a stopping criterion.
Our experiments use the \href{https://pypi.org/project/missingpy/}{missingpy}
package to perform MissForest imputation.

\begin{figure}[H]
    \includegraphics[width=0.8\textwidth]{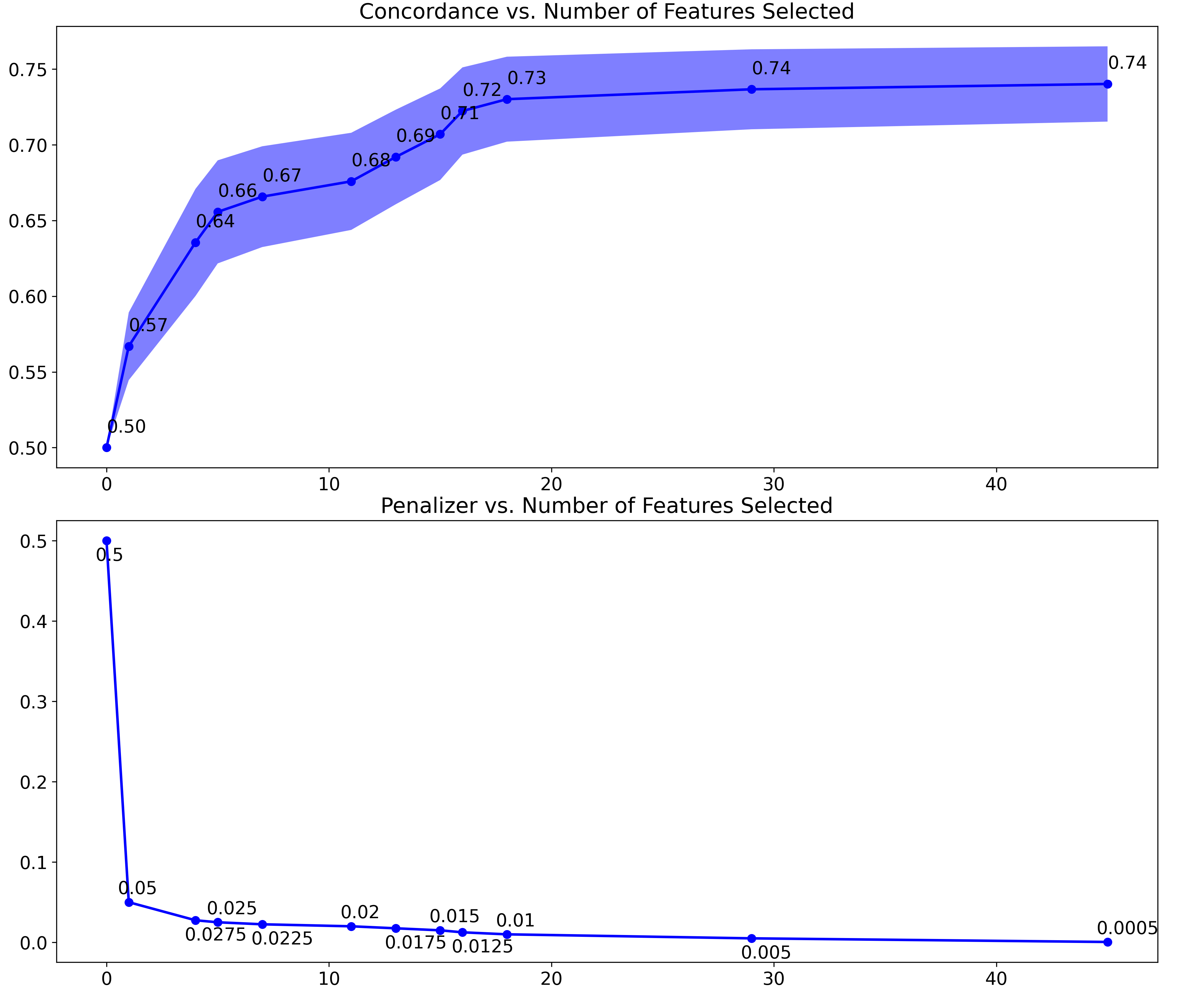} 
    \caption{Tradeoff from controlling the penalty hyperparameter $\lambda$ in Lasso-Cox. As $\lambda$ decreases, more features are selected and concordance increases. Beyond $\lambda = 0.01$, the gain in performance levels off.}
    \label{fig:missforest10_concordance}
\end{figure}

To check the sensitivity of model selection to MissForest imputation,
we re-ran the imputation algorithm with 10 different seeds 
and re-ran grid search to find the best set of parameters on each of these 10 datasets.
As discussed in the paper, Lasso-Cox with a penalty ($\lambda$) level of 0.01 
consistently achieved high concordance (around 0.73) while only selecting approximately 18 same features out of a total of 52 features across the 10 different seeds. 
Selecting additional features by lowering the penalty level 
did not yield substantial gains in performance 
(Figure \ref{fig:missforest10_concordance}). 

\begin{figure}[H]
    \centering
    \includegraphics[width=\textwidth]{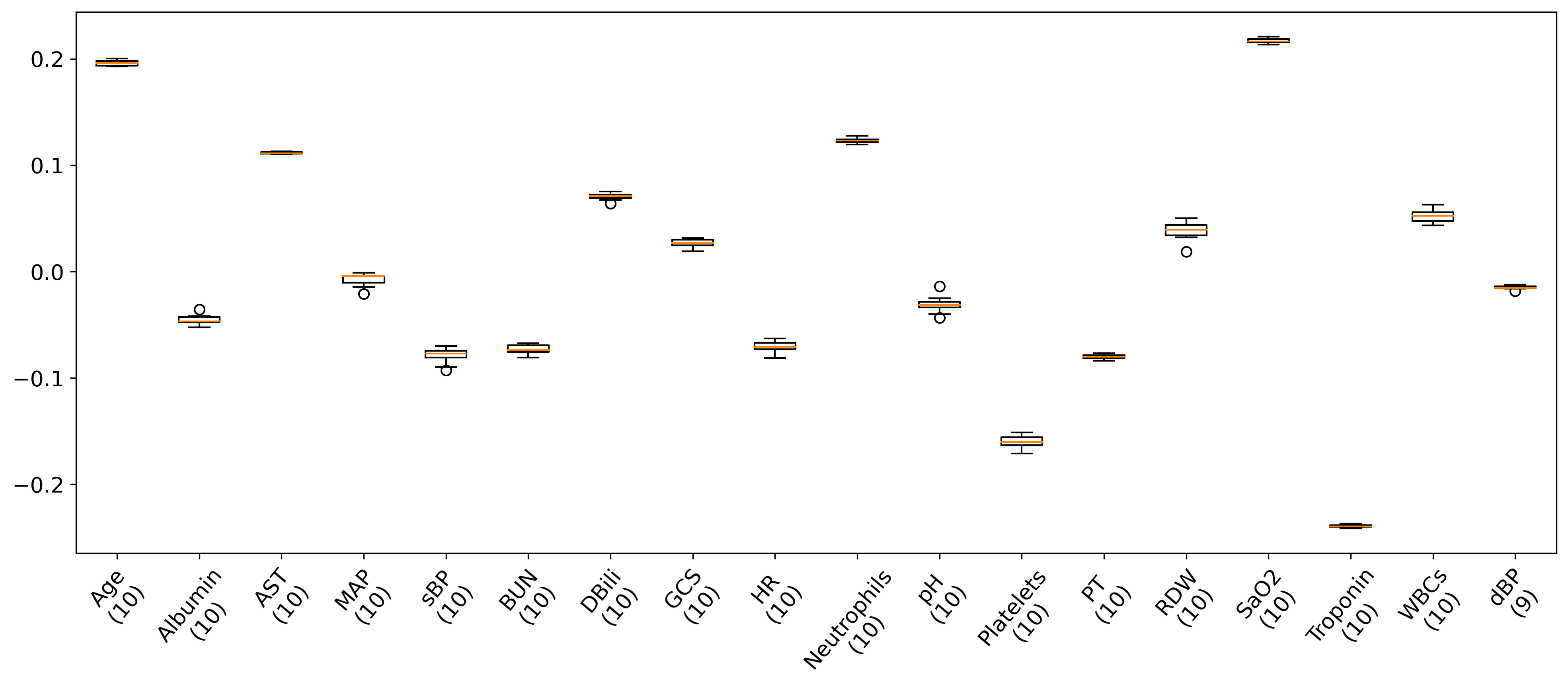}
    \caption{Coefficients learned by the L1-regularized cox model with penalty 0.01 from ten random forest imputations of the data. The x-axis has the variable name (number of times that the variable has nonzero coefficient across 10 runs)}
    \label{fig:missforest10_coefficients}
\end{figure}

With a penalty level of 0.01, Lasso-Cox was consistent 
in the features selected and coefficients it learned. 
Figure \ref{fig:missforest10_coefficients} shows that 
across all except one of the ten random forest imputations of the dataset
(in which it selected one less feature), 
all 18 features were selected and the model 
learned relatively consistent coefficients. 

\newpage
\section{Survival Curves for High and Low Risk Groups} \label{appendix:survcurves}
\begin{figure}[H]
    \centering
    \includegraphics[width=0.7\textwidth]{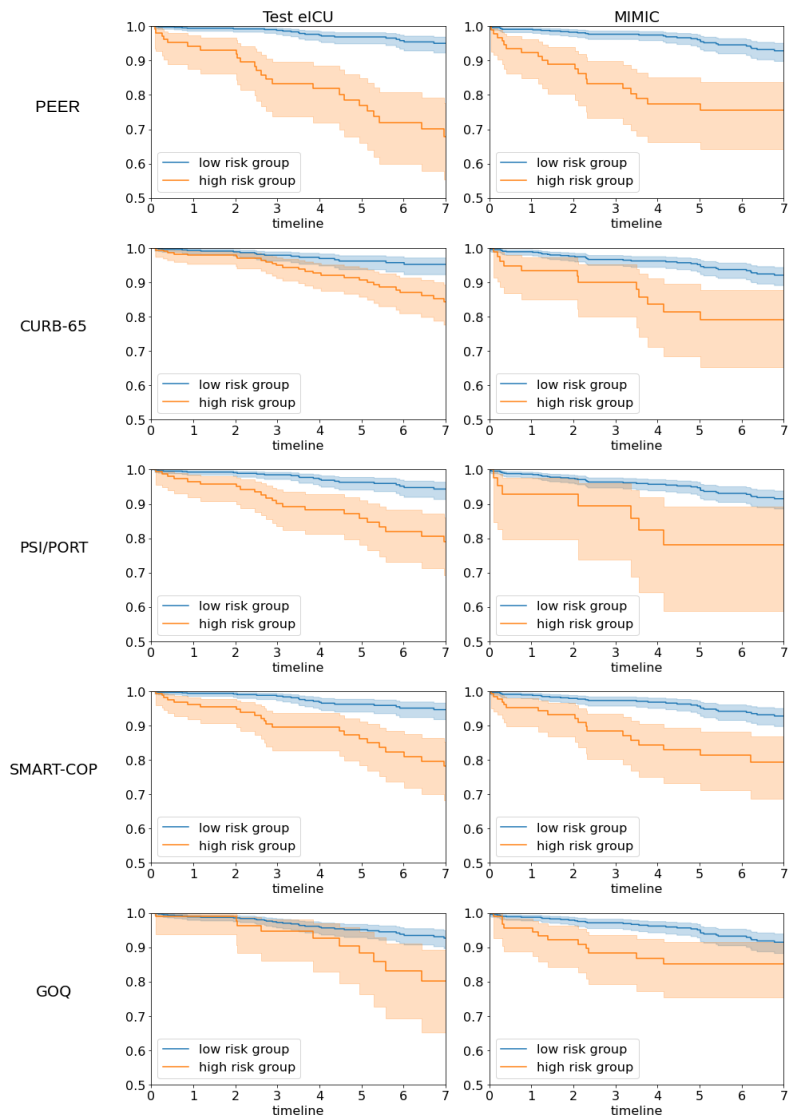}
    \caption{Comparison of survival curves corresponding to risk strata derived from the PEER score, CURB-65, PSI/PORT, SMART-COP, and GOQ}
    \label{fig:allscores_strata}
\end{figure}

Figure \ref{fig:allscores_strata} contains survival curves 
for high and low risk groups derived from the PEER score (our model), 
CURB-65, PSI/PORT, SMART-COP, and GOQ.
The high/low risk groups for CURB-65 and PSI/PORT 
were created by following the criteria listed in their respective papers 
where former one to greater or equal to 3 and the later one to higher or equal to 130. 
For SMART-COP and GOQ, since we did not find a cutoff criterion in their papers,
we just follow our criterion to set the cutoff point 
as the 90th percentitle of the train set risk. 
\end{document}

%% file: tables/cohort_table.tex
\begin{tabular}{lll}
\hline
                                          Variable &           eICU (n = 3617)&          MIMIC (n = 937) \\
\hline
                             \textbf{Demographics} &                      &                      \\
                                        Age, years &     58.0 (48.0-64.0) &     54.5 (44.1-62.7) \\
                                  Age range, years &                      &                      \\
                                \hspace{5mm}18-30 &          225 (6.2\%) &           83 (8.9\%) \\
                                 \hspace{5mm}30-39 &          277 (7.7\%) &          94 (10.0\%) \\
                                 \hspace{5mm}40-49 &         500 (13.8\%) &         159 (17.0\%) \\
                                 \hspace{5mm}50-59 &        1064 (29.4\%) &         281 (30.0\%) \\
                             \hspace{5mm}60-70 &        1546 (42.7\%) &         320 (34.2\%) \\
                                            Gender &                      &                      \\
                                  \hspace{5mm}Male &        1949 (53.9\%) &         542 (57.8\%) \\
                                \hspace{5mm}Female &        1663 (46.0\%) &         395 (42.2\%) \vspace{0.5em}\\
                   \textbf{Physical exam findings} &                      &                      \\
                                       Orientation &                      &                      \\
                              \hspace{5mm}oriented &        1121 (31.0\%) &         411 (43.9\%) \\
                              \hspace{5mm}confused &        1287 (35.6\%) &           76 (8.1\%) \\
                           Temperature (\degree C) &     36.9 (36.6-37.3) &     37.2 (36.6-37.7) \\
                     Heart rate (beats per minute) &    89.0 (77.0-101.0) &    90.0 (78.0-104.0) \\
             Respiratory rate (breaths per minute) &     20.0 (17.0-25.0) &     20.0 (16.0-25.0) \\
                    Systolic blood pressure (mmHg) &  120.0 (106.0-136.0) &  118.0 (104.0-134.0) \\
                   Diastolic blood pressure (mmHg) &     66.0 (57.0-76.0) &     63.0 (54.0-72.0) \\
                     Mean arterial pressure (mmHg) &     81.0 (72.0-93.0) &     79.0 (71.0-90.0) \\
                                Glasgow Coma Scale &     14.0 (10.0-15.0) &      14.0 (9.0-15.0) \vspace{0.5em}\\
                      \textbf{Laboratory findings} &                      &                      \\
                                        Hemotology &                      &                      \\
     \hspace{5mm}Red blood cells (millions/$\mu$L) &        3.5 (3.0-4.0) &        3.4 (3.0-3.8) \\
  \hspace{5mm}White blood cells (thousands/$\mu$L) &      11.0 (7.9-15.6) &      11.0 (8.0-15.1) \\
          \hspace{5mm}Platelets (thousands/$\mu$L) &  193.0 (136.0-261.0) &  199.0 (128.8-276.0) \\
                       \hspace{5mm}Hematocrit (\%) &     31.1 (27.2-35.6) &     30.2 (27.0-33.6) \\
       \hspace{5mm}Red blood cell dist. width (\%) &     15.2 (14.0-16.8) &     14.8 (13.8-16.4) \\
          \hspace{5mm}Mean corpuscular volume (fL) &     90.4 (86.0-95.0) &     89.0 (85.0-93.0) \\
 \hspace{5mm}Mean corpuscular hemoglobin/ MCH (pg) &     29.7 (27.9-31.2) &     30.2 (28.7-31.6) \\
              \hspace{5mm}MCH concentration (g/dL) &     32.7 (31.7-33.6) &     33.8 (32.8-34.8) \\
                      \hspace{5mm}Neutrophils (\%) &     82.0 (73.3-89.0) &     82.3 (73.8-88.5) \\
                      \hspace{5mm}Lymphocytes (\%) &       8.4 (5.0-14.0) &       9.5 (5.8-15.7) \\
                        \hspace{5mm}Monocytes (\%) &        6.0 (3.7-8.6) &        4.0 (2.7-5.9) \\
                      \hspace{5mm}Eosinophils (\%) &        0.1 (0.0-1.0) &        0.4 (0.0-1.2) \\
                        \hspace{5mm}Basophils (\%) &        0.0 (0.0-0.3) &        0.1 (0.0-0.3) \\
                       \hspace{5mm}Band cells (\%) &       8.0 (3.0-17.0) &        0.0 (0.0-5.0) \\
                                         Chemistry &                      &                      \\
                       \hspace{5mm}Sodium (mmol/L) &  139.0 (136.0-142.0) &  139.0 (136.0-142.0) \\
                    \hspace{5mm}Potassium (mmol/L) &        3.9 (3.6-4.3) &        3.9 (3.6-4.3) \\
                     \hspace{5mm}Chloride (mmol/L) &  105.0 (101.0-109.0) &  105.0 (101.0-109.0) \\
                  \hspace{5mm}Bicarbonate (mmol/L) &     25.0 (22.0-28.0) &     26.0 (23.0-29.0) \\
           \hspace{5mm}Blood urea nitrogen (mg/dL) &     19.0 (12.0-33.0) &     17.0 (11.0-28.0) \\
                    \hspace{5mm}Creatinine (mg/dL) &        0.8 (0.6-1.4) &        0.8 (0.6-1.3) \\
                       \hspace{5mm}Glucose (mg/dL) &  131.0 (105.0-165.0) &  124.0 (104.5-151.5) \\
  \hspace{5mm}Aspartate aminotransferase (units/L) &     30.0 (19.0-57.0) &     37.0 (22.0-70.0) \\
    \hspace{5mm}Alanine aminotransferase (units/L) &     27.0 (16.0-47.0) &     28.0 (18.0-52.0) \\
        \hspace{5mm}Alkaline phosphatase (units/L) &    84.0 (62.0-117.0) &    85.0 (62.0-121.0) \\
             \hspace{5mm}C-reactive protein (mg/L) &      19.6 (8.8-42.2) &      35.9 (8.3-98.8) \\
               \hspace{5mm}Direct bilirubin (mg/L) &        0.2 (0.1-0.5) &        0.6 (0.2-2.2) \\
                \hspace{5mm}Total bilirubin (mg/L) &        0.5 (0.3-0.8) &        0.6 (0.4-1.1) \\
                  \hspace{5mm}Total protein (g/dL) &        6.0 (5.3-6.7) &        6.1 (5.3-7.0) \\
                       \hspace{5mm}Calcium (mg/dL) &        8.2 (7.7-8.6) &        8.2 (7.8-8.6) \\
                        \hspace{5mm}Albumin (g/dL) &        2.6 (2.2-3.1) &        3.0 (2.6-3.5) \\
                      \hspace{5mm}Troponin (ng/mL) &        0.1 (0.0-0.2) &        0.0 (0.0-0.3) \\
                                       Coagulation &                      &                      \\
                \hspace{5mm}Prothrombin time (sec) &     14.5 (12.7-16.7) &     13.9 (13.0-15.3) \\
     \hspace{5mm}Partial thromboplastin time (sec) &     33.0 (28.5-41.0) &     30.2 (26.6-36.9) \\
                                         Blood gas &                      &                      \\
                                    \hspace{5mm}pH &     7.39 (7.33-7.43) &     7.41 (7.36-7.45) \\
     \hspace{5mm}Partial pressure of oxygen (mmHg) &    83.0 (68.0-111.0) &    97.0 (73.5-127.5) \\
     \hspace{5mm}Arterial oxygen saturation (mmHg) &     96.0 (94.0-99.0) &     97.0 (95.0-98.0) \vspace{0.5em}\\
                             \textbf{Outcomes} &                      &                      \\
                                          Deceased &          270 (7.5\%) &          94 (10.0\%) \\
                         Vasopressors administered &         589 (16.3\%) &         389 (41.5\%) \\
                                   Ventilator used &        1835 (50.7\%) &         758 (80.9\%) \\
\hline
\end{tabular}

%% file: tables/betas.tex
\begin{tabular}{lc}
\hline
                              Feature &            HR (95\% CI) \\
\hline
 Age (years) &  1.22 (1.04 -- 1.43) \\
 Heart rate (beats per minute) &  1.13 (0.984 -- 1.3) \\
 Systolic blood pressure (mmHg) &  0.928 (0.755 -- 1.14) \\
 Diastolic blood pressure (mmHg) &  0.996 (0.745 -- 1.33) \\
 Mean arterial pressure (mmHg) &  0.926 (0.673 -- 1.27) \\
 Glasgow Coma Scale &  0.93 (0.803 -- 1.08) \\
 White blood cells (thousands/$\mu$L) &  0.984 (0.871 -- 1.11) \\
 Platelets (thousands/$\mu$L) &  0.924 (0.79 -- 1.08) \\
 Red blood cell dist. width (\%) &  1.24 (1.08 -- 1.43) \\
 Neutrophils (\%) &  0.972 (0.853 -- 1.11) \\
 Blood urea nitrogen (mg/dL) &  1.07 (0.937 -- 1.23) \\
 Aspartate aminotransferase (units/L) &  1.12 (1.06 -- 1.18) \\
 Direct bilirubin (mg/L) &  1.03 (0.935 -- 1.13) \\
 Albumin (g/dL) &  0.954 (0.82 -- 1.11) \\
 Troponin (ng/mL) &  1.06 (0.985 -- 1.14) \\
 Prothrombin time (sec) &  1.05 (0.909 -- 1.2) \\
 pH &  0.856 (0.75 -- 0.977) \\
 Arterial oxygen saturation (mmHg) &  0.787 (0.723 -- 0.856) \\

\hline
\end{tabular}

%% file: tables/concordances.tex
\begin{tabular}{llll}
\hline
      Score &          Train eICU &          Test eICU &        MIMIC \\
      & concordance & concordance & concordance\\
\hline
 PEER (ours)&  \textbf{0.77 (0.72 - 0.81)} &  \textbf{0.77 (0.69 - 0.83)} &  \textbf{0.66 (0.57 - 0.74)} \\
 CURB-65~\citep{curb65}&  0.66 (0.61 - 0.70) &  0.62 (0.55 - 0.69) &  0.59 (0.52 - 0.66) \\
 PSI/PORT~\citep{psiport}&  0.71 (0.66 - 0.76) &  0.71 (0.63 - 0.78) &  0.62 (0.55 - 0.69) \\
 SMART-COP~\citep{smartcop}&  0.69 (0.64 - 0.73) &  0.73 (0.67 - 0.80) &  \textbf{0.66 (0.59 - 0.72)} \\
 GOQ~\citep{maastrichtE6}&  0.67 (0.63 - 0.71) &  0.62 (0.54 - 0.70) &  0.58 (0.50 - 0.66) \\
\hline
\end{tabular}

%% file: tables/missingness.tex
\begin{tabular}{lll}
\hline
                              Variable & eICU (n = 3617) & MIMIC (n = 937) \\
\hline
                                   Age &             0.001 (5) &               0.0 (0) \\
                                Gender &             0.001 (5) &               0.0 (0) \\
                      Pleural effusion &               0.0 (0) &               0.0 (0) \\
                           Orientation &          0.334 (1209) &            0.48 (450) \\
               Temperature (\degree C) &            0.006 (20) &           0.182 (171) \\
         Heart rate (beats per minute) &            0.009 (32) &            0.018 (17) \\
 Respiratory rate (breaths per minute) &             0.001 (3) &            0.017 (16) \\
        Systolic blood pressure (mmHg) &           0.063 (229) &            0.023 (22) \\
       Diastolic blood pressure (mmHg) &           0.063 (229) &            0.023 (22) \\
         Mean arterial pressure (mmHg) &           0.079 (287) &            0.018 (17) \\
                    Glasgow Coma Scale &            0.27 (977) &            0.016 (15) \\
     Red blood cells (millions/$\mu$L) &            0.012 (44) &              0.01 (9) \\
  White blood cells (thousands/$\mu$L) &            0.006 (22) &             0.009 (8) \\
          Platelets (thousands/$\mu$L) &            0.014 (49) &              0.01 (9) \\
                        Hematocrit (\%) &            0.006 (23) &              0.01 (9) \\
        Red blood cell dist. width (\%) &           0.057 (207) &            0.012 (11) \\
          Mean corpuscular volume (fL) &            0.025 (91) &            0.011 (10) \\
 Mean corpuscular hemoglobin/ MCH (pg) &           0.074 (269) &            0.011 (10) \\
              MCH concentration (g/dL) &            0.025 (91) &              0.01 (9) \\
                       Neutrophils (\%) &            0.24 (869) &           0.152 (142) \\
                       Lymphocytes (\%) &           0.167 (603) &            0.15 (141) \\
                         Monocytes (\%) &           0.177 (641) &           0.152 (142) \\
                       Eosinophils (\%) &           0.209 (755) &           0.152 (142) \\
                         Basophils (\%) &           0.256 (925) &           0.152 (142) \\
                        Band cells (\%) &          0.752 (2720) &           0.454 (425) \\
                       Sodium (mmol/L) &            0.004 (13) &              0.01 (9) \\
                    Potassium (mmol/L) &            0.008 (29) &             0.009 (8) \\
                     Chloride (mmol/L) &            0.009 (33) &             0.009 (8) \\
                  Bicarbonate (mmol/L) &           0.057 (207) &             0.009 (8) \\
           Blood urea nitrogen (mg/dL) &            0.004 (14) &              0.01 (9) \\
                    Creatinine (mg/dL) &            0.007 (27) &              0.01 (9) \\
                       Glucose (mg/dL) &            0.006 (23) &            0.011 (10) \\
  Aspartate aminotransferase (units/L) &           0.174 (628) &           0.218 (204) \\
    Alanine aminotransferase (units/L) &           0.177 (640) &           0.219 (205) \\
        Alkaline phosphatase (units/L) &           0.184 (665) &           0.223 (209) \\
             C-reactive protein (mg/L) &          0.946 (3420) &           0.916 (858) \\
               Direct bilirubin (mg/L) &          0.808 (2923) &            0.82 (768) \\
                Total bilirubin (mg/L) &           0.185 (670) &           0.227 (213) \\
                  Total protein (g/dL) &           0.184 (664) &           0.841 (788) \\
                       Calcium (mg/dL) &            0.021 (77) &            0.027 (25) \\
                        Albumin (g/dL) &            0.16 (577) &           0.279 (261) \\
                      Troponin (ng/mL) &          0.591 (2138) &           0.505 (473) \\
                Prothrombin time (sec) &          0.396 (1431) &            0.035 (33) \\
     Partial thromboplastin time (sec) &          0.545 (1973) &            0.038 (36) \\
                                    pH &           0.244 (883) &            0.104 (97) \\
     Partial pressure of oxygen (mmHg) &           0.223 (807) &           0.134 (126) \\
     Arterial oxygen saturation (mmHg) &            0.015 (54) &           0.726 (680) \\
                              Deceased &               0.0 (0) &               0.0 (0) \\
             Vasopressors administered &               0.0 (0) &               0.0 (0) \\
                       Ventilator used &               0.0 (0) &               0.0 (0) \\
\hline
\end{tabular}

%% file: tables/mean_std.tex
\begin{tabular}{lll}
\hline
          Feature &   mean & std. dev. \\
\hline
 wbc &  12.9 &  8.91 \\
 platelets &  208 &  108 \\
 rdw &  15.8 &  2.47 \\
 neutrophils &  79.1 &  13 \\
 bun &  25.1 &  19.5 \\
 ph &  7.38 &  0.0713 \\
 direct\_bilirubin &  0.385 &  0.816 \\
 albumin &  2.65 &  0.636 \\
 pt &  16.6 &  6.75 \\
 ast &  143 &  774 \\
 troponin &  1.07 &  3.85 \\
 age &  54.4 &  12.5 \\
 heart\_rate &  89.4 &  17.8 \\
 sao2 &  95.8 &  4.12 \\
 gcs &  11.3 &  3.26 \\
 bp\_systolic &  122 &  22 \\
 bp\_diastolic &  67.7 &  15.1 \\
 bp\_mean\_arterial &  83.7 &  17.9 \\
\hline
\end{tabular}